\documentclass[twocolumn,showpacs,superscriptaddress, secnumarabic,amssymb, nobibnotes, aps, prd]{revtex4}

\usepackage{graphicx}
\usepackage{dcolumn}
\usepackage{bm}
\usepackage{amsmath,amsthm}
\usepackage{amsfonts}
\usepackage{epstopdf}
\usepackage[caption=false]{subfig}
\begin{document}

\title{Determination of Intrinsic Magnetic Response from Local Measurements of Fringing Fields}%
\author{B. Wen}
\affiliation{Department of Physics, City College of New York, CUNY, New York, New York
10031, USA}
\author{A.J. Millis}
\affiliation{Department of Physics, Columbia University, New York, New York 10027, USA}
\author{E. Pardo}
\affiliation{Institute of Electrical Engineering, Slovak Accademy of Sciences, Dubravska 9, 841 04 Bratislava, Slovakia}
\author{P. Subedi   }
\affiliation{Department of Physics, New York University, New York, New York 10003, USA}
\author{A. D. Kent}
\affiliation{Department of Physics, New York University, New York, New York 10003, USA}
\author{Y. Yeshurun}
\affiliation{Department of Physics, Institute of Nanotechnology, Bar-Ilan University,
Ramat-Gan 52900, Israel}
\author{M.P. Sarachik}
\affiliation{Department of Physics, City College of New York, CUNY, New York, New York
10031, USA}

\begin{abstract}
Micron-sized Hall bars and micro-SQUIDs are now used routinely to measure the local static and dynamic magnetic response with micron-scale spatial resolution.  While this provides a powerful new tool, determining the intrinsic magnetization presents new challenges, as it requires correcting for demagnetization fields that vary widely with position on a sample.  In this paper we develop a method to correct for the demagnetization effect at local points of a rectangular prism shaped sample using a finite element analysis of Maxwell's equation applied to local Hall sensor measurements calibrated by bulk measurements of the magnetization.  This method can be generalized to other geometric shapes to analyze data obtained with local magnetic probes.
\end{abstract}
\maketitle
\section{Introduction}

The demagnetizing field is the magnetic field $H_d$ generated by the magnetization in a material.  For a paramagnet, it is related to an externally applied field $H_a$ (taken here to be spatially uniform), the measured magnetization $M(\mathbf{r})$, and the magnetic susceptibility $\chi$, via
\begin{equation}
\mathbf M = \chi(\mathbf{H_a}+\mathbf{H_d}).
\end{equation}

The demagnetizing field is directed opposite to the magnetization, and for a magnetization measurement taken on a bulk macroscopic sample, its magnitude is generally approximated by a single {\it demagnetization factor} $N$, so that $\mathbf H_d = - N\cdot \mathbf{M}$.  The demagnetizing factor can be calculated analytically only for ellipsoids of revolution, which have uniform magnetization.  For some specimens of simple shape, the demagnetizing factor is calculated by empirical formulas; tabulated values are available for particular shapes, including cylinders \cite{ChenIEEE1991}, square bars \cite{PardoIEEE2004} and rectangular prisms \cite{PardoJAP2002,PardoIEEE2005} .

However, local measurements of magnetization measure a combination of the local field and the local demagnetization field, both of which may  vary substantially from point to point, so that one cannot use a single global demagnetization factor $N$.   In other words:
\begin{equation}
\mathbf H_d(\mathbf{r}) = - {\underline{\underline N}}(\mathbf{r}) \mathbf{M}(\mathbf{r}).
\end{equation}

\begin{figure}[htbp]
\centering
\includegraphics[width=0.6\linewidth]{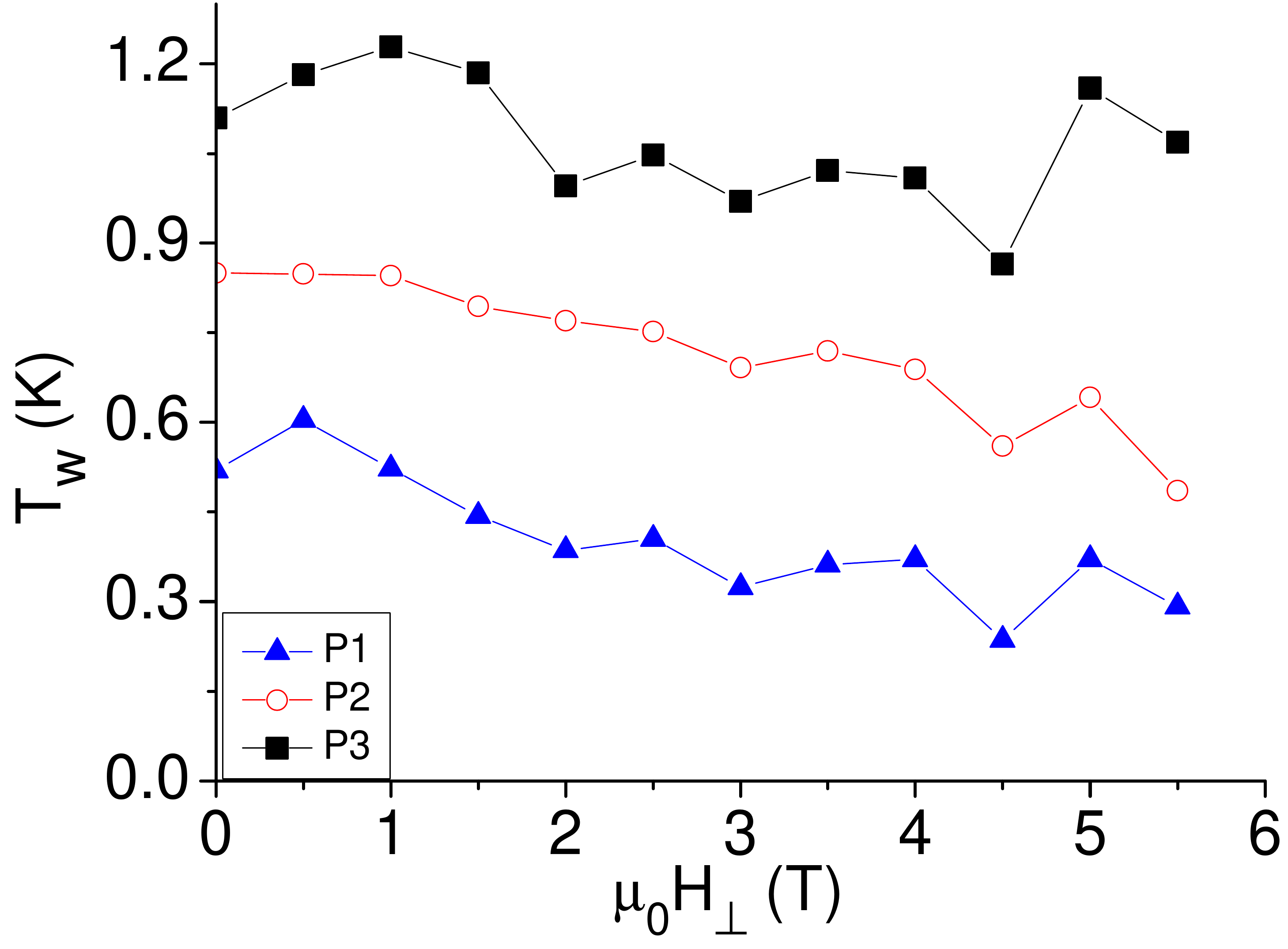}
\caption{The Weiss temperature $T_W$ deduced from micron-sized Hall sensor measurements taken at three different points on a rectangular prism-shaped sample of the molecular magnet Mn$_{12}$-acetate is shown as a function of magnetic field applied transverse to the easy axis.  For each Hall sensor, $T_W$ was extrapolated from plots of the inverse susceptibility versus temperature; see Ref. \cite{SubediPRB2012} for details and Appendix \ref{critiqueAppendix} for discussion.  Measured relative to one end of the $1.85$ mm-long sample, P1 is at $0$ mm, P2 is at $0.21$ mm, P3 is at $0.43$ mm.  Except for Fig. \ref{zeroField}, all the experimental data discussed and shown in this paper refer to this particular sample.}
\label{TcwComparison}
\end{figure}

A correct interpretation of local measurements enabled by local sensors such as micron-scale Hall and micro-SQUIDs therefore requires a full nonlocal magnetostatic analysis of the demagnetization effects. The need was revealed by recent local measurements obtained by micron-sized Hall sensors placed on the surface of millimeter-sized crystals of the prototypical molecular magnet Mn$_{12}$-acetate \cite{SubediPRB2012}. These sensors in effect measure the normal component of the magnetic field H at the sample surface at the sensor position. Although the temperature- and field-dependence were similar, measurements on the same sample taken by Hall sensors at different positions yielded widely different values for the magnetization and thermodynamic quantities derived from these measurements, such as the Weiss temperature, as demonstrated in Fig. \ref{TcwComparison}. This was traced to the fact that the demagnetizing field depends crucially on  the placement of the Hall bar on the sample as well as the sample shape.

In this paper we develop a model for interpreting micro-Hall bar magnetometry data and demonstrate its applicability for measurements on millimeter sized rectangular prism shaped samples of Mn$_{12}$-acetate. A crucial finding is that the nonlinearity implicit in the magnetostatic equation (magnetization produces demagnetizing field which in turn produces magnetization) has a significant effect on the micro-Hall bar measurements, most particularly as one approaches a magnetic ordering temperature where the susceptibility becomes very large.

\section{Theory}\label{theory}
\begin{figure}[htbp]
\centering
\includegraphics[width=0.6\linewidth]{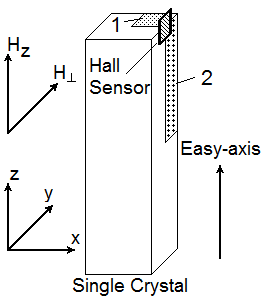}
\caption{Schematic of the sample and Hall sensor.  The rectangular prism-shaped sample has boundaries $-x_0<x<x_0$, $-y_0<y<y_0$, $-z_0<z<z_0$. The Hall sensor (labeled as such) is on the surface $x=x_0$ centered at ($x_0$, $y_1$, $z_1$), and covers the area $(y_1-a)<y<(y_1+a)$, $(z_1-b)<z<(z_1+b)$.  Calculated results shown in Fig.~\ref{sigma} (a) refer to the area labeled 1; calculated results shown in Fig.~\ref{sigma} (b) and Fig.~\ref{Sur} refer to area 2.}
\label{setupSingle}
\end{figure}

The physical situation of interest can be approximated by an idealized model in which the sample is taken to be a perfect rectangular prism as depicted in Fig.~\ref{setupSingle} oriented so that the applied magnetic field is parallel to four of the faces.  A Hall sensor placed on a face measures the component of the magnetic  field perpendicular to the face; the design of the sensors is such that the field component is in effect measured at the surface of the sample \cite{BoThesis}. As shown in Fig. \ref{setupSingle}, the applied field is parallel to the plane of the Hall sensor in the experimentally relevant case, and the field component measured by the Hall sensor arises only from demagnetizing fields.

The demagnetizing fields are computed from the standard magnetostatic equations relating the magnetic field $B$, the magnetic induction $H$ and the magnetization $M$
\begin{eqnarray}
{\mathbf \nabla}\cdot {\mathbf B}&=&0
\label{nablaB}
\\
\mathbf{B}&=&\mu_0  (\mathbf{H}+\mathbf{M})
\label{BH}
\end{eqnarray}
We assume that the magnetization is non-zero only inside the sample volume defined above and is related to the magnetic induction by a  susceptibility tensor  ${\underline{\underline \chi }}$ which is local and the same at all points in the sample, so that
\begin{equation}
\mathbf{M}(\mathbf{r})={\underline{\underline \chi }}\mathbf H(\mathbf r) \;\;\mathrm{}
\label{UniformChiEq}
\end{equation}
inside the sample.  This is a good approximation so long as one can ignore defects and sample degradation that could cause variations in the local susceptibility.

In this paper we will deal with insulators (no free currents) so that the magnetic induction can be represented as the gradient of a  magnetic potential $\Phi_M$ as
\begin{equation}
\mathbf{H}=-\triangledown\Phi_M.
\label{PhiM}
\end{equation}
Then from Eq.~\ref{nablaB} we conclude
\begin{eqnarray}
\triangledown^2\Phi_M&=&\triangledown\cdot\mathbf{M}
\label{Phieqn}
\end{eqnarray}
where $\mathbf{M}\neq0$ only inside the sample.

The general solution to Eq.~\ref{Phieqn}, is \cite{jackson}:
\begin{equation}
\Phi_M(\mathbf{r})=-\frac{1}{4\pi}\int \frac{\triangledown'\cdot\mathbf{M}(\mathbf{r}')}{|\mathbf{r}-\mathbf{r}'|}d^3r'\label{PhiMsolution}
\end{equation}
where $\mathbf{M}(\mathbf{r}^\prime)$ is the limit of $\mathbf{M}$ as it approaches the boundary point $\mathbf{r}^\prime$ from within the sample.

We next integrate the right hand side of Eq.~\ref{PhiMsolution} by parts and define the unit normal at position ${\mathbf r^\prime}$ to be ${\mathbf e_n^\prime}$ obtaining
\begin{eqnarray}
\nonumber  \Phi_M({\mathbf r})=-\frac{1}{4\pi}\oint_A \frac{{\mathbf e_n^\prime} \cdot {\mathbf M}({\mathbf r}^\prime)}{\left|{\mathbf r}-{\mathbf r}^\prime\right|}d^2r^\prime\\
  +\frac{1}{4\pi}\int {\mathbf \nabla}^\prime\left( \frac{1}{\left|{\mathbf r}-{\mathbf r}^\prime\right|}\right)\cdot {\mathbf M}({\mathbf r}^\prime)d^3r^\prime
\label{PhiMintegral1}
\end{eqnarray}
Moving the derivative in the second term from the $r^\prime$ to the $r$, taking the gradient of $\Phi_M$, noting that $\nabla^2\frac{1}{|r-r^\prime|}\sim \delta^3(r-r^\prime)$, evaluating the result for $r$ {\em outside} the sample volume and adding the applied field $\mathbf{H}_a$ (assumed spatially uniform) gives
\begin{equation}
{\mathbf H}({\mathbf r})={\mathbf H}_a+\frac{1}{4\pi}\oint \frac{\left({\mathbf r}-{\mathbf r}^\prime\right)}{\left|{\mathbf r}-{\mathbf r}^\prime\right|^3}\sigma({\mathbf r}^\prime)d^2r^\prime
\label{Hfinal}
\end{equation}
with
\begin{equation}
\sigma({\mathbf r})={\mathbf e_n} \cdot {\mathbf M}({\mathbf r})
\label{sigmadef}
\end{equation}

To transform Eq.~\ref{Hfinal} to a form amenable to numerical analysis we first take the limit as $\mathbf{r}$ approaches the sample surface from outside, and then consider only the component of Eq.~\ref{Hfinal} corresponding to fields normal to the sample surface. This gives
\begin{equation}
\mathbf{e}_n\cdot\mathbf{H}(\mathbf{r})-\frac{1}{4\pi}\oint \frac{\left({\mathbf r}-{\mathbf r}^\prime\right)}{\left|{\mathbf r}-{\mathbf r}^\prime\right|^3}\sigma({\mathbf r}^\prime)d^2r^\prime=\mathbf{e}_n\cdot\mathbf{H}_a
\label{Hfinal1}
\end{equation}
with $\mathbf{r}$ now assumed to lie on the sample surface. We further assume (as is the case in the experimental situation of interest here) that the sample is a rectangular parallelepiped with planar surfaces parallel to the principle axes of the susceptibility tensor $\chi$ which we take to be independent of position within the material.  Because $\mathbf{H}(\mathbf{r})$ is continuous across the interface we then can write
\begin{equation}
\mathbf e_n \cdot\mathbf{H}(\mathbf{r})=\frac{1}{\chi_n}\sigma(\mathbf{r})
\label{chiinv}
\end{equation}
with $\chi_n$ the eigenvalue of the susceptibility tensor relevant to fields perpendicular to the surface at point $\mathbf{r}$.

Combining Eqs.~\ref{Hfinal1}, \ref{chiinv} we finally obtain
\begin{equation}
\oint_\mathcal{A}d^2r^\prime \mathcal{D}(\mathbf{r},\mathbf{r^\prime})\sigma(\mathbf{r^\prime})=\mathbf{e}_n\cdot\mathbf{H}_a
\label{final}
\end{equation}
with $\mathcal{A}$ the surface of the solid and
\begin{equation}
\mathcal{D}(\mathbf{r},\mathbf{r^\prime})=\frac{1}{\chi_n}\delta^2(\mathbf{r}-\mathbf{r^\prime})-\frac{1}{4\pi} \frac{\left({\mathbf r}-{\mathbf r}^\prime\right)}{\left|{\mathbf r}-{\mathbf r}^\prime\right|^3}
\label{Ddef}
\end{equation}

Eq.~\ref{final} can be solved by standard finite element methods. However, before discussing this analysis we discuss some general features of the solution. We are interested in the case that the applied field $\mathbf H_a$  is parallel to four of the sample surfaces (``the sides'') and perpendicular to the other two (``top'' and ``bottom'') (see Fig. \ref{setupSingle}), while the Hall bar is placed on the sample sides and detects the field component normal to the side and thus also perpendicular to $\mathbf H_a$. With this in mind we now iteratively solve Eq.~\ref{final} assuming the susceptibility is small. To leading ($-1$) order in $\chi^{-1}$, $\mathcal{D}(\mathbf{r},\mathbf{r^\prime})=\frac{1}{\chi_n}\delta^2(\mathbf{r}-\mathbf{r^\prime})$ so $\sigma(\mathbf{r})=0$ for $\mathbf{r}$ on the sides and $\sigma(\mathbf{r})=\mp \chi_{zz}\mathbf H_a$ for $\mathbf{r}$ on the top ($-$) or bottom ($+$) of the sample. To this order the fringing fields vanish and a Hall sensor mounted on the side of the sample would measure nothing.

The leading contribution to the field measured by a Hall sensor mounted on the side of the sample is obtained by iterating the equation, writing $\sigma=\sigma^{(0)}+\sigma^{(1)}$ with $\sigma^{(0)}=\mp \chi_{zz}\mathbf H_a$ for $\mathbf{r}$ on the top ($+$) or bottom ($-$) of the sample and zero otherwise, and $\sigma^{(1)}$ to be determined. Inserting this into Eq.~\ref{final} and collecting terms of order $\chi$ we obtain
\begin{equation}
\sigma^{(1)}(\mathbf{r})-\frac{1}{4\pi}\oint \frac{\left({\mathbf r}-{\mathbf r}^\prime\right)}{\left|{\mathbf r}-{\mathbf r}^\prime\right|^3}\sigma^{(0)}(\mathbf{r^\prime})=0
\label{firstorder}
\end{equation}
Using the explicit form of $\sigma^{(0)}$ and converting from $\mathbf M$ to $\mathbf H$ gives, for $\mathbf{r}$ on the side of the sample
\begin{equation}
\mathbf e_n\cdot\mathbf{H}(\mathbf{r})=\chi_{zz}\mathbf H_a\left(\oint_{bottom}-\oint_{top}\right)d^2r^\prime\mathcal{D}(\mathbf{r},\mathbf{r^\prime})
\label{secondorder}
\end{equation}
Thus we see that for small values of the susceptibility the value of the field measured by a Hall bar sensor mounted on the side of the sample is proportional to the applied field, to the component of the susceptibility parallel to the applied field, and to the difference of the propagator integrated over the top and bottom of the sample. If the susceptibility is not too large, the measured micro-Hall bar signal is proportional to the magnetic susceptibility but with a coefficient that varies with position, in particular vanishing at high symmetry points, in our case half way between the two planes. However, as $\chi$ increases, the second order approximation used to derive Eq.~\ref{secondorder} becomes inadequate. We observe that in general $\mathcal{D}$ has negative eigenvalues, so the demagnetization  corrections act to reduce the magnitude of the magnetization induced by an applied  field; we also observe that the reduction will not be uniform along the sample surface. Thus, the degree to which demagnetization effects distort the measurement
is a complicated function of the magnitude of $\chi$ and of the position of the micro-Hall sensor. Our previous experiments showed that these effects are important in practice. To understand them in detail, a numerical
solution of Eq. \ref{final} is required.

We use the finite-element method of Pardo \emph{et al.} \cite{ChenIEEE1991,PardoIEEE2004,PardoIEEE2005} to numerically solve Eq.~\ref{final}, which is briefly summarized as follows. We cover the surface with $N$ non-overlapping rectangular tiles. Their size is chosen to keep the increment of pole density in each direction being roughly uniform. Thus the area $\mathcal A_i$ of tile $i$ is smaller on the edges and corners where the magnetization varies strongly in space, and larger in the middle of the sample where the magnetization changes slowly. The functional dependence of the size in each direction $\delta x_i$, $\delta y_i$ and $\delta z_i$ on the coordinate of the tiles' center position is described in Ref. \cite{PardoJAP2002}.

The number of divisions in each direction, $2n_x$, $2n_y$, and ($2n_z+1$) (with a layer riding on the $z=0$ midplane with $\sigma=0$), is taken by fixing the number of elements integrally belonging to the $x$, $y$, $z>0$ region. The elements centered at $z=0$ have $\sigma^i=0$, so that $\sigma^i$ on these elements are known variables. For the sample we presented in this paper, $a=b=0.3$ mm, $c=1.85$ mm, so we set $n_x=n_y$, and we assume $n_z/n_x$ is approximately $c/a$. We tested three cases: (a) $n_x=n_y=10$, $n_z=55$, (b) $n_x=n_y=13$, $n_z=74$ and (c) $n_x=n_y=20$, $n_z=119$. The results differ by less than $0.04\%$; we conclude that the element sizes are small enough to yield reliable results. In the study that follows, we have assumed case (a).

We define a matrix $\underline{\underline D}$ (see Appendix \ref{finiteElementAppendix} for details) with components
\begin{equation}
D_{ij}=\frac{1}{4\pi A_i}\int_{A_i}(\int_{A_j}\frac{\mathbf{e}_n(\mathbf{r}_i)\cdot\left(\mathbf{r}_i-\mathbf{r}_j\right)}{|\mathbf{r}_i-\mathbf{r}_j|^3}d^2r_j)d^2r_i
\label{Ddef}
\end{equation}
and approximate the solution of  Eq.~\ref{final} as
\begin{equation}
\chi_n^{-1}\sigma(\mathbf{r})=\sum_j\left(\delta_{ij}-\chi_n D_{ij}\right)^{-1}\mathbf e^j_n\cdot \mathbf H_a \label{sigmaeq}
\end{equation}
For materials such as Mn$_{12}$ that have anisotropic susceptibilities, see Appendix \ref{anisotropyAppendix} for additional information.

\section{Results}\label{results}
We have solved Eq. \ref{sigmaeq} for a range of representative cases, using typical experimental geometries and typical values of susceptibility ($\chi_{zz} \sim 0.1 - 1$). Below we showed a case of $\chi_{xx}=\chi_{yy}=0.016, \chi_{zz}=0.40$.
\begin{figure}[htbp]
\centering
\subfloat[$\sigma$ on the $z=z_0$ surface (the end of the crystal).]{
\label{sigma:1}
\includegraphics[width=0.45\linewidth]{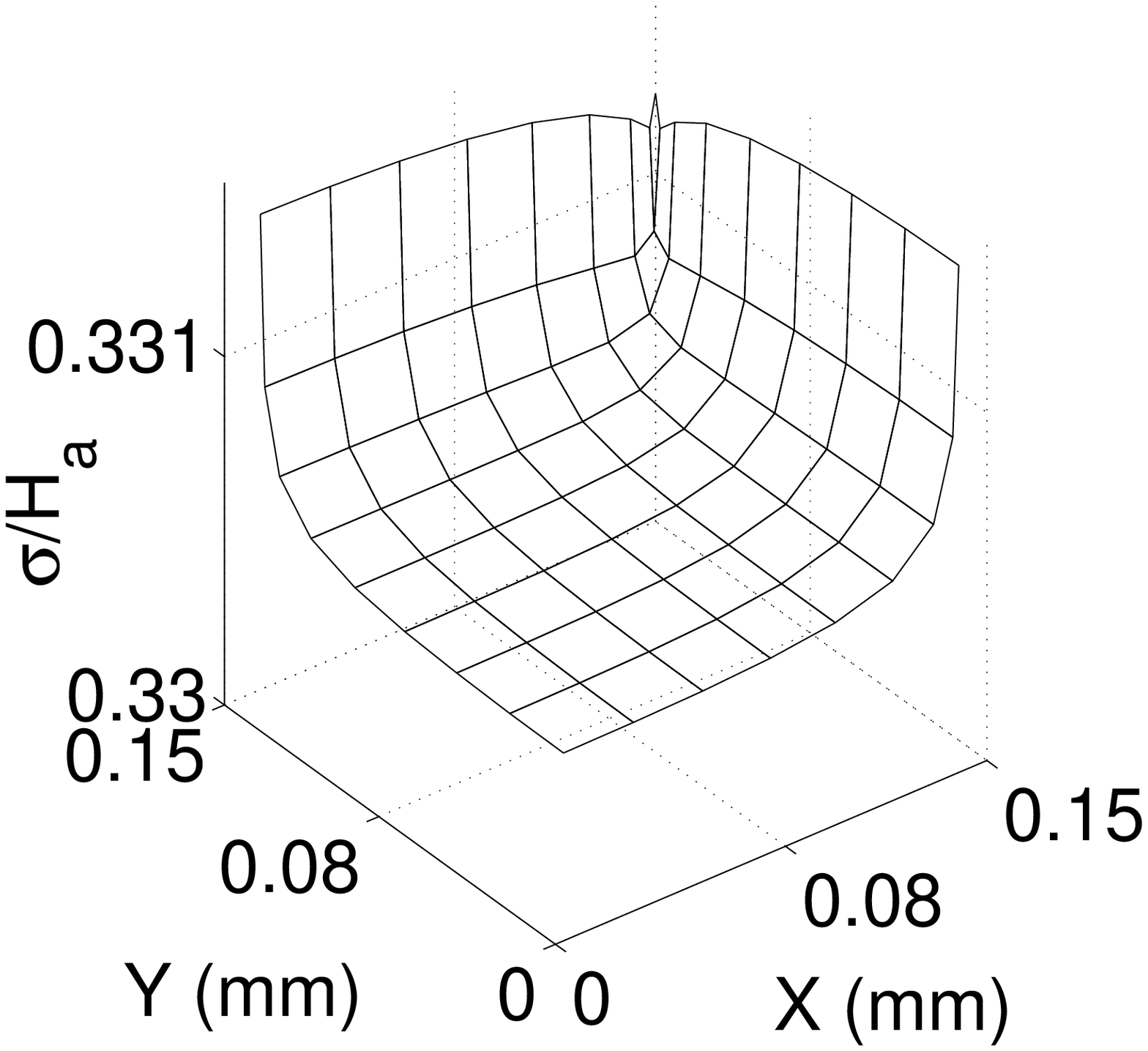}}
\hspace{0.01\linewidth}
\subfloat[$\sigma$ on the $x=x_0$ surface (the side of the crystal).]{
\label{sigma:2}
\includegraphics[width=0.45\linewidth]{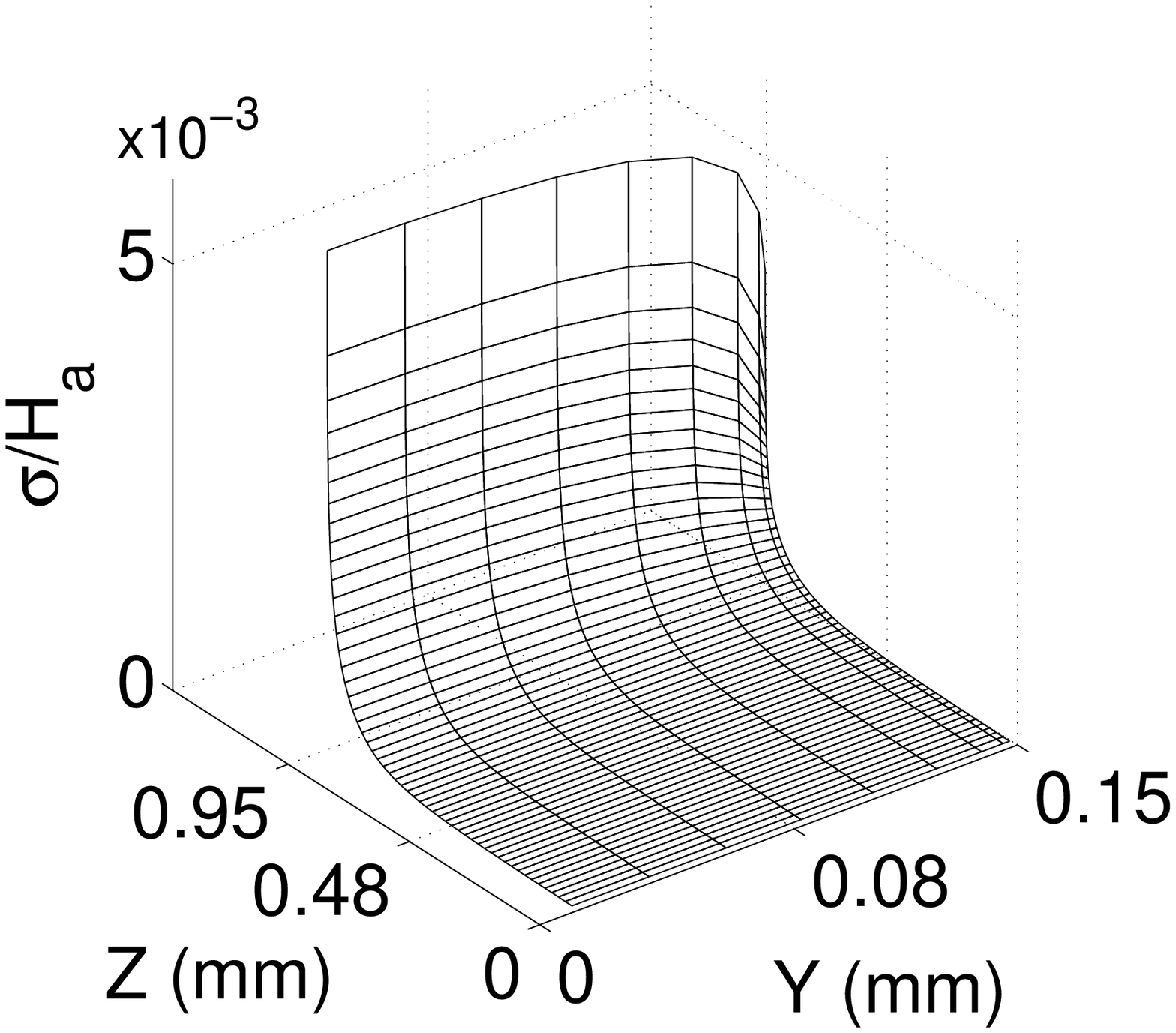}}
\caption{An example of the calculated distribution of the induced surface magnetic pole density $\sigma$ normalized by the value of applied magnetic field $H_a$ on a rectangular prism-shaped crystal of Mn$_{12}$-acetate. The coordinates are defined in Fig. \ref{setupSingle} with $x_0=y_0=0.15$ mm and $z_0=0.95$ mm.  Frame (a) and (b) refer to area 1 and 2, respectively, shown in Fig.~\ref{setupSingle} In this calculation we set $H_\perp=0$ T, $H_z=0.1$ T. The unit of $\sigma / H_a$ is dimensionless.}
\label{sigma}
\end{figure}

Figure \ref{sigma} shows the surface pole distribution $\sigma$  induced  by a field applied along the z axis. As expected, the surface poles are largest on the top and bottom surfaces (left panel, surface normal parallel to the applied field); the demagnetizing field weakens near the edges so the moments become larger.  The moments are smaller on the side panels, and exhibit the spatial variation qualitatively expected from the small $\chi$ analysis: the field is largest near the top and bottom, and vanishes at the midpoint.

\begin{figure}[htbp]
\centering
\subfloat[$H_x$]{
\label{Sur:1}
\includegraphics[width=0.45\linewidth]{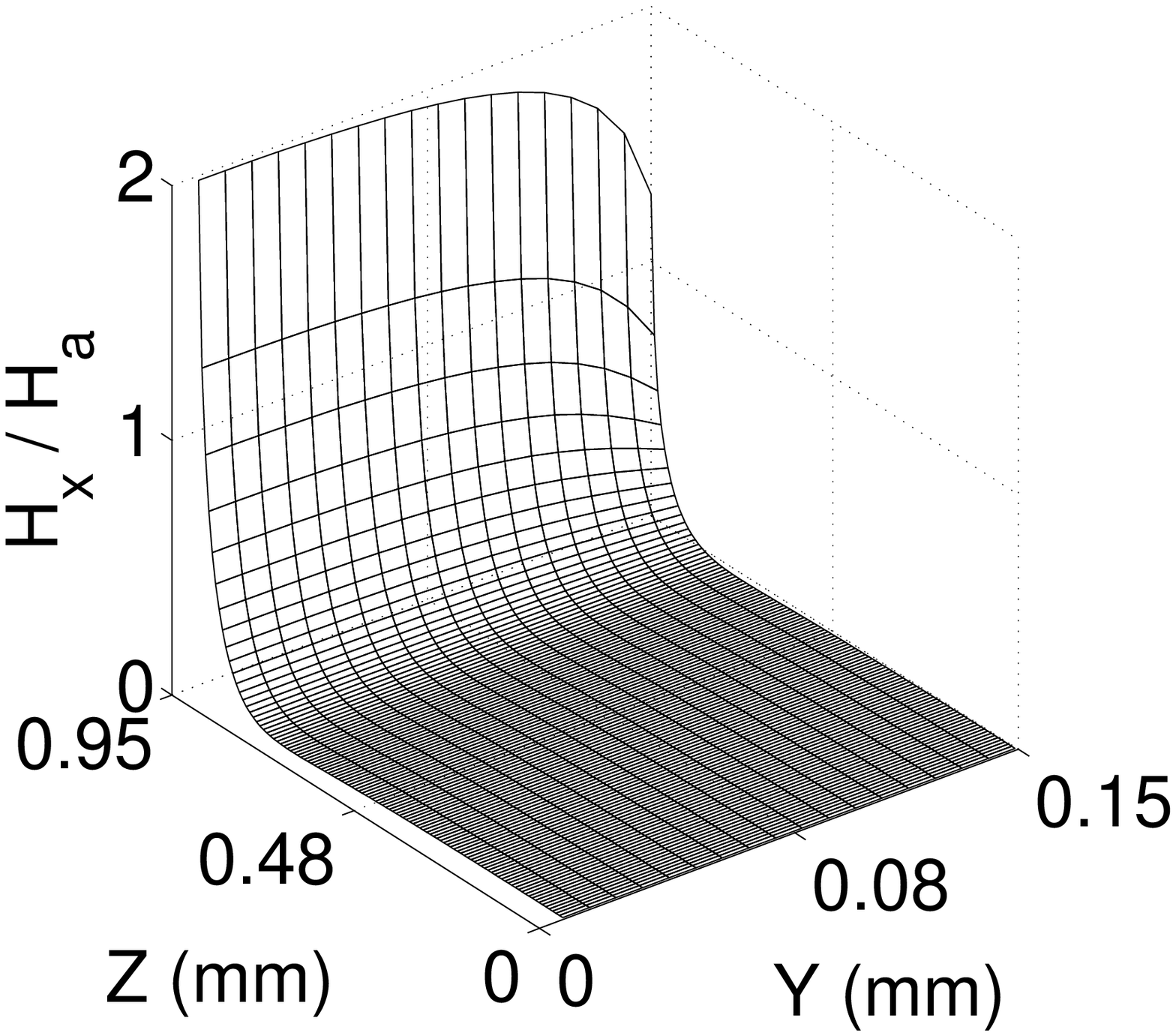}}
\hspace{0.01\linewidth}
\subfloat[$H_y$]{
\label{Sur:2}
\includegraphics[width=0.45\linewidth]{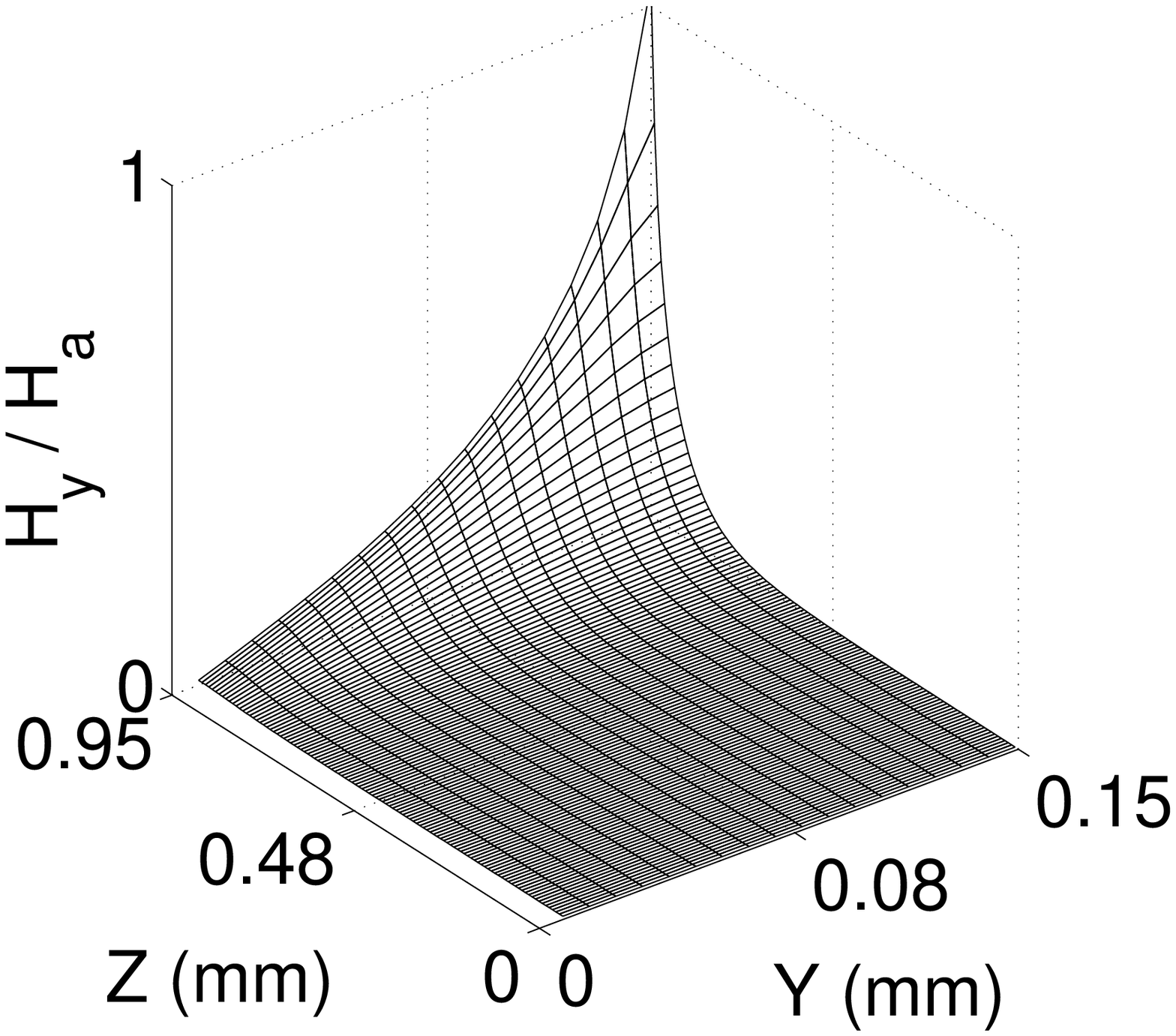}}
\hspace{0.01\linewidth}
\subfloat[$H_z$]{
\label{Sur:3}
\includegraphics[width=0.45\linewidth]{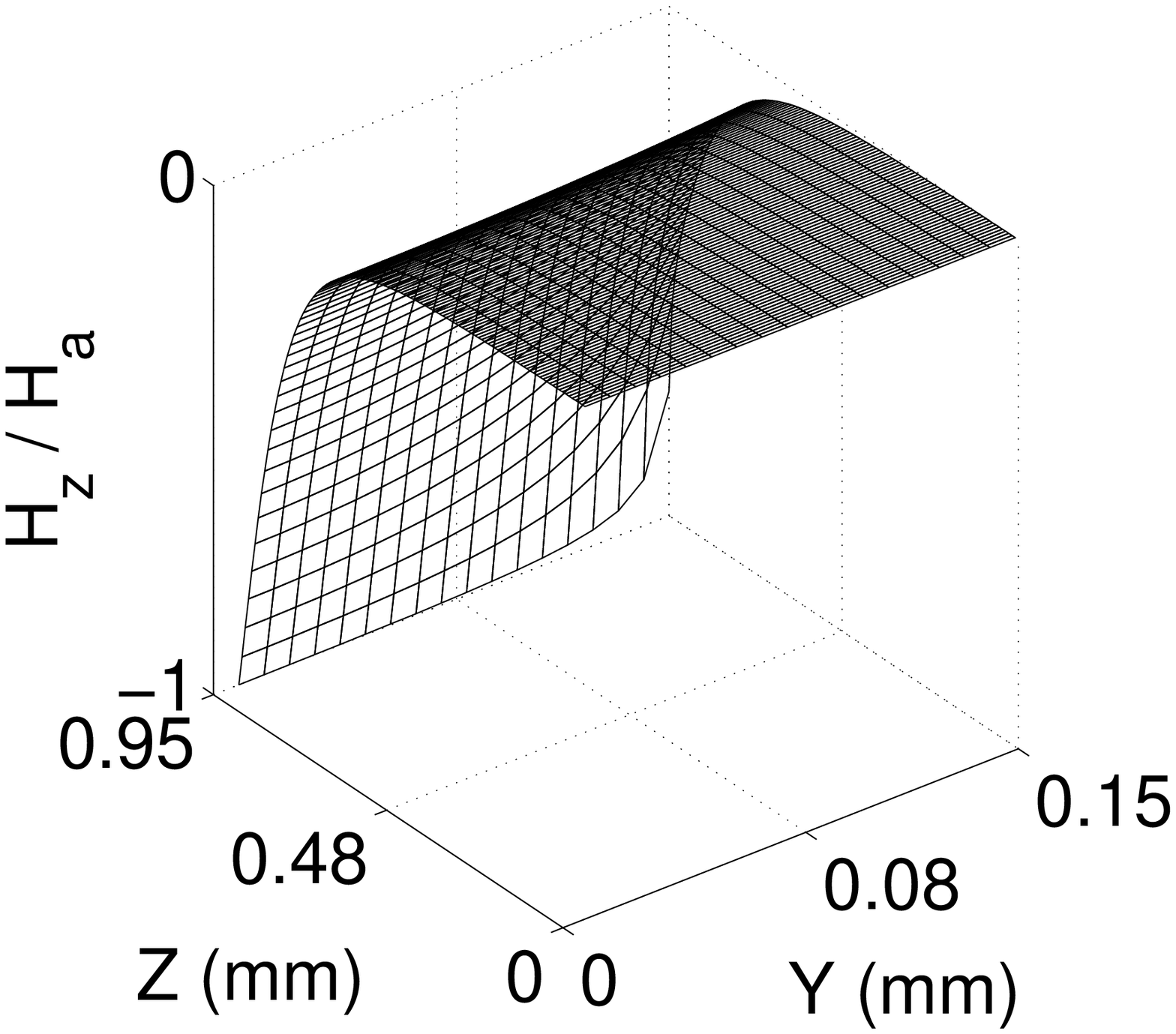}}
\caption{The magnetic field on the $x=x_0$ surface of the crystal normalized by the applied magnetic field $H_a$ (area 2 in Fig.~\ref{setupSingle}). The coordinates are defined in Fig. \ref{setupSingle} with $x_0=y_0=0.15$ mm and $z_0=0.95$ mm.}
\label{Sur}
\end{figure}

Figure \ref{Sur} shows the induced magnetic field in three directions on the $x=x_0$ surface of the crystal. The field in the $x$ direction is bigger near the end of the crystal; the field in the $y$ direction is almost zero except near the corner close to the $y=y_0$ surface; the field in the $z$ direction is negative, which means the field direction is opposite to the externally applied field, as expected for a demagnetization field. Using this, we can calculate the corresponding Hall sensor signal and compare with the experimental result. We define $H_{x,\mathrm{Hall}}(\mathbf r)=<H_x(\mathbf r)>$, integrating the $x$ component of the field (Fig. \ref{Sur:1}) over the area where the Hall sensor is positioned to obtain the expected Hall sensor signal:
\begin{align}
V_\mathrm{Hall-calculated}(\mathbf r)=\alpha\cdot H_{x,\mathrm{Hall}}(\mathbf r), \label{Hall_effect}
\end{align}
where the Hall effect coefficient $\alpha$ is obtained experimentally by calibrating the Hall sensor response as a function of magnetic field applied perpendicular to the surface.

\begin{figure}[htbp]
\centering
\includegraphics[width=0.7\linewidth]{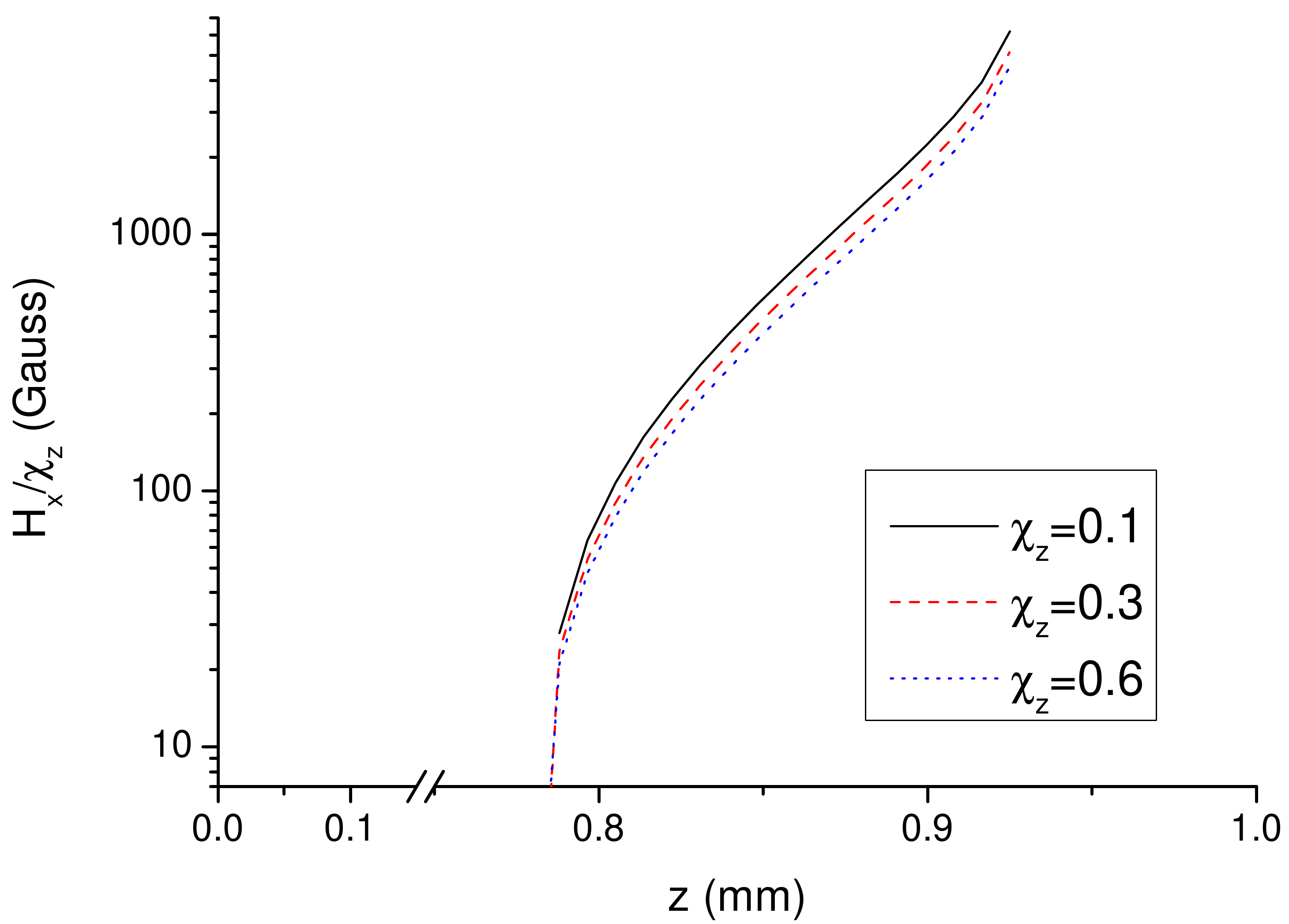}
\caption{Calculated results for $\mathbf H_{x}|_{y=0}(z)$ for different values of $\chi_{zz}$; the values of $\mathbf H_{x}$ are normalized by the corresponding $\chi_{zz}$.}
\label{hxvsz}
\end{figure}

Figure \ref{hxvsz} shows the calculated induced $H_x$ as a function of distance away from the sample center along the middle line ($y=0$) of the $x=x_0$ surface for several values of the susceptibility within the expected range for Mn$_{12}$-ac between 3  and 15 K. It is clear that as $\chi_{zz}$ gets bigger, the field near the edge increases.

In sections \ref{theory} and \ref{results}, we presented a method for extracting the $V_\mathrm{Hall}$ for a given intrinsic susceptibility $\chi$ in an external magnetic field as well as the inverse process, namely, extracting $\chi$ from $V_\mathrm{Hall}$. Both cases require that geometric factors such as aspect ratio and sensor location be specified. In the next section, we summarize these steps, and apply the method to analyze experimental data \cite{SubediPRB2012}.

\section{APPLICATION OF METHOD TO EXPERIMENT}
In Reference \cite{SubediPRB2012} we presented a method to obtain the demagnetization-corrected susceptibility $\chi$ for samples of M$_{12}$-acetate as a function of temperature in the absence of a transverse magnetic field. Using this result as a starting point, we now extend the method to correct for the demagnetization factor for measurements taken in the presence of a transverse field, i.e., obtaining the intrinsic $\chi$ from the apparent $\chi_\mathrm{app-meas}=H_{x,\mathrm{Hall,measured}}/H_a=V_\mathrm{Hall-measured}/\alpha/H_a$.

Using the numerical method in section \ref{results}, we are able to calculate $V_\mathrm{Hall-calculated}$ using a given values of $\chi$ for the given sample aspect ratio. We then obtain the calculated apparent susceptibility $\chi_\mathrm{app-cal}=H_{x,\mathrm{Hall,calculated}}/H_a=V_\mathrm{Hall-calculated}/\alpha/H_a$ from Eq. \ref{Hall_effect}. This process can be summarized by expressing the Hall sensor signal as a function $\zeta$ of the intrinsic susceptibility:
\begin{align}
\chi_\mathrm{app-cal}(T,H_\perp)=\zeta_{\{\mathbf r, c/a, H_\perp, \alpha\}}(\chi(T,H_\perp))\label{hall-cal}
\end{align}
where $\zeta$ is a function of the parameters $\mathbf r$ (the Hall sensor location), $c/a$ (aspect ratio of the sample), $H_\perp$ (applied transverse field) and $\alpha$ (Hall effect coefficient of the sensor).

\begin{figure}[htbp]
\centering
\includegraphics[width=0.7\linewidth]{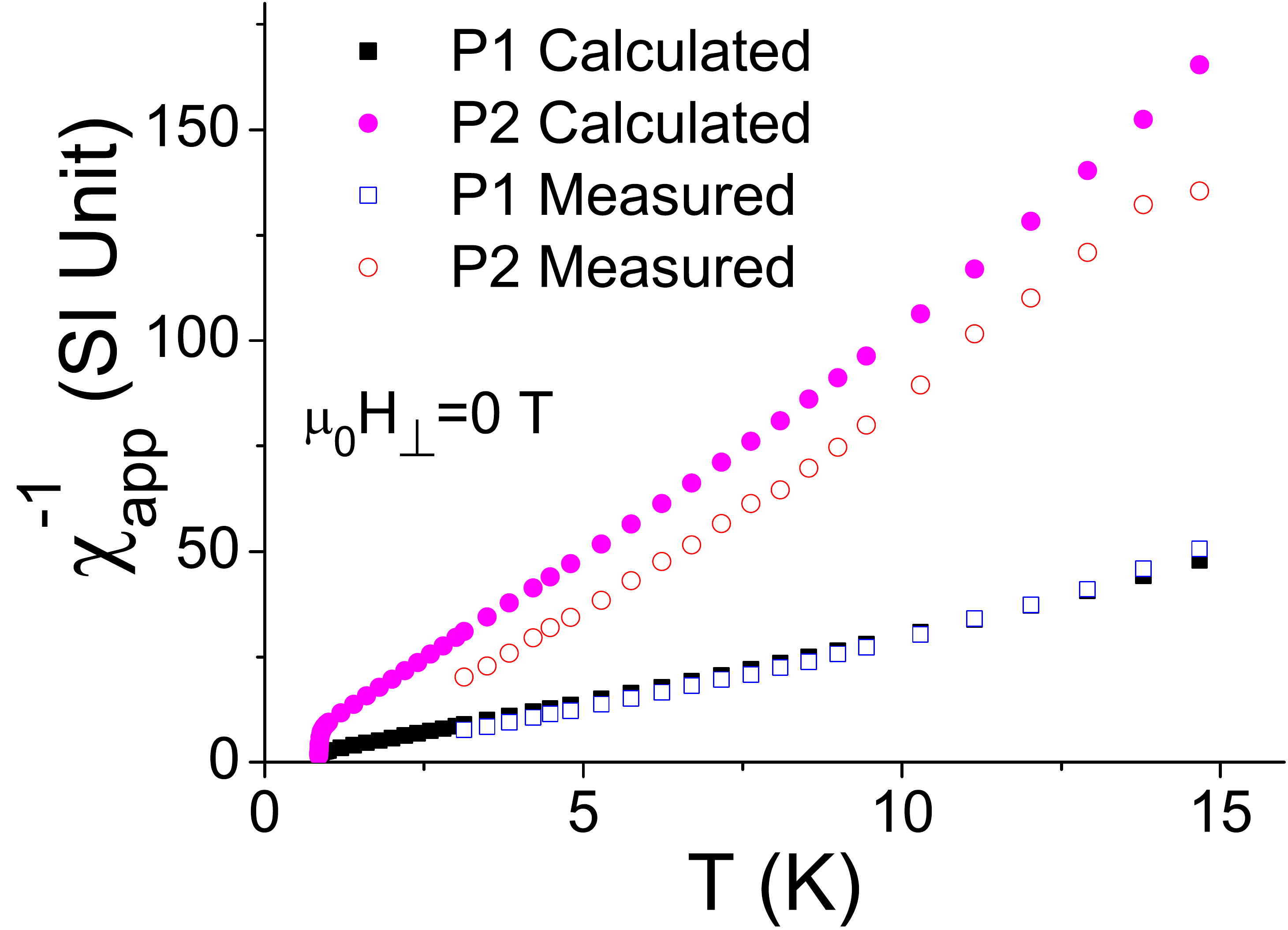}
\caption{Temperature dependence of the calculated (closed symbols) and measured (open symbols) inverse apparent susceptibility of Mn$_{12}$-ac for zero transverse field. P1 and P2 are two different sensor locations as defined in Fig. \ref{TcwComparison}.}
\label{chiinv_compare}
\end{figure}

Figure \ref{chiinv_compare} shows a comparison between the a direct Hall sensor measured $\chi_\mathrm{app-meas}$ and the calculated $\chi_\mathrm{app-cal}$ for a same sample using the above method. The value of intrinsic $\chi(T,0)$ used in the calculation is deduced from measurements (in a Quantum Design MPMS magnetometer) of bulk samples of M$_{12}$-acetate as a function of aspect ratio in the limit of infinitely long, thin samples for which the demagnetization corrections are negligible \cite{ShiqiPRB2010}.  As shown in this figure, for different sensor locations, the calculated results differ from the measurements by different amounts. In addition to a multiplicative factor, which is simply a calibration of the signal amplitude, an additional shift in the temperature axis is required to collapse the lines onto a single curve.

Both $\chi_\mathrm{app-meas}^{-1}$ and $\chi_\mathrm{app-cal}^{-1}$ are linear between $3$ K and $6$ K for zero transverse field up to $H_\perp=6$ T, and can be denoted as $\chi_\mathrm{app-cal}^{-1}=a_1\cdot (T-T_1)$ and $\chi_\mathrm{app-meas}^{-1}=a_2\cdot (T-T_2)$. We can eliminate $T$ in the equations to get $\chi^{-1}_\mathrm{app-cal}=a_1\cdot (\chi^{-1}_\mathrm{app-meas}/a_2+T_2-T_1)$.

The difference between the $\chi_\mathrm{app-cal}$ and $\chi_\mathrm{app-meas}$ originates from several factors, including the uncertainty in the measured coefficient $\alpha$ in Eq. \ref{Hall_effect}, the uncertainty of the Hall-sensor active area, deviations from the assumed sample shape from a perfect rectangular prism, and other possible sources.

We assumed that none of the factors are affected by a transverse magnetic field, i.e., the fitting constant $a_1$, $T_1$ and $a_2$, $T_2$ are determined by factors other than $H_\perp$. They can be calibrated at zero field, where the $\chi_\mathrm{app-cal}$ can be calculated using the demagnetization corrected $\chi$ obtained from MPMS measurements. This process can be abstracted as follows:
\begin{align}
\left.\begin{array}{r}
\chi(T,0)\xrightarrow{Eq. \ref{hall-cal} }\chi_\mathrm{app-cal}(T,0)\nonumber\\
\chi_\mathrm{app-meas}(T,0)
\end{array}\right\}\xrightarrow{} a_1, T_1; a_2, T_2
\end{align}

In order to interpret our results in the presence of a transverse field,  $H_\perp$, we now apply the same horizontal shift and multiplication to the $\chi_\mathrm{app-meas}(T)$ as was done in zero transverse field to obtain $\chi_\mathrm{app-cal}$ in the presence of a constant transverse field, $H_\perp\neq0$. Then we use relationship Eq. \ref{hall-cal} to deduce the $\chi(T)$ from the inferred $\chi_\mathrm{app-cal}$. This process can be abstracted as follows:
\begin{align}
&\chi_\mathrm{app-meas}(T,H_\perp)\xrightarrow{\chi^{-1}_\mathrm{app-cal}=a_1\cdot (\chi^{-1}_\mathrm{app-meas}/a2+T_2-T_1)} \nonumber\\
&\chi_\mathrm{app-cal}(T,H_\perp)\xrightarrow{Eq. \ref{hall-cal} }\chi(T,H_\perp) \label{chiRelation}
\end{align}

\begin{figure}[htbp]
\centering
\hspace{-0.1in}
\includegraphics[width=1.0\linewidth]{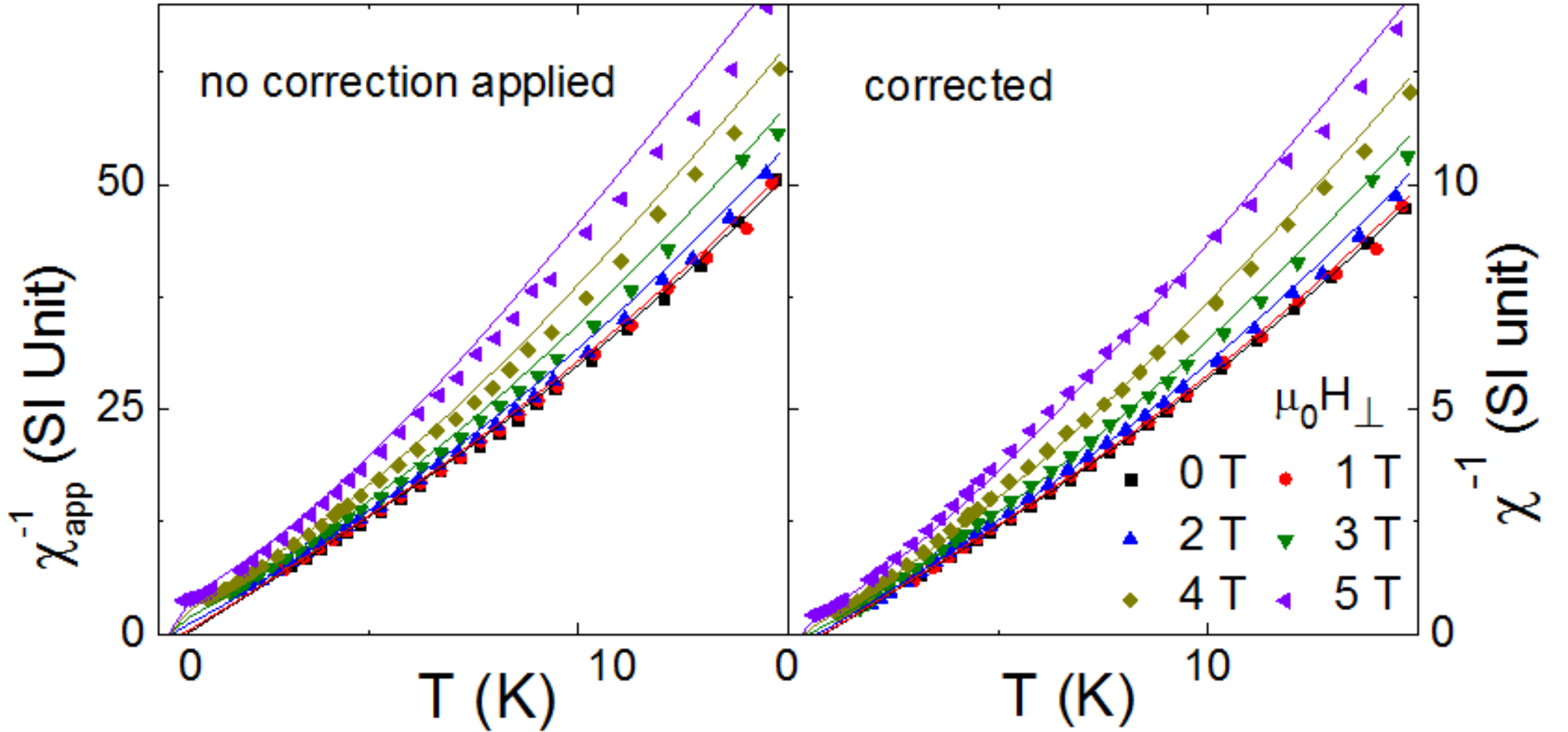}
\caption{Susceptibility $\chi$ versus temperature in different transverse magnetic field for the P1 position on the Mn$_{12}$-ac sample.  In both panels, the lines are the $\chi_{MFA}$ results obtained by a mean field calculation of Millis {\it et al.} \cite{MillisPRB2010}.  The symbols data (a) as taken, and (b) corrected by the method described above.} \label{allField}
\end{figure}

In Fig. \ref{allField} we compare the resulting $\chi(T)$ with a mean field approximation (MFA) $\chi_{MFA}(T)$ calculation using the Random Field Ising Ferromagnet (RFIFM) model described in Ref. \cite{MillisPRB2010, SubediPRB2012}. Good agreement is obtained, providing validation of the assumptions made in our calculations for the demagnetization correction.

\section{Summary}
We have shown that micron-scale magnetization measurements require an analysis to account for the effect of demagnetizing fields that goes well beyond the usual linear approximation.  As shown in this paper, the measured field is linearly proportional to the applied magnetic field within a restricted range of the parameters, while strongly nonlinear behavior obtains for large values of the susceptibility, most particularly in the regions near the ends of a sample where the Hall bar signal is largest.

We summarize the proposed steps that need to be taken to correct for the demagnetization effect when using local micron-scale Hall sensors.

First, for a given sample geometry and sensor placement (in our case, a right rectangular prism with Hall bars placed along one face), one must calculate the relationship between the intrinsic susceptibility and the measured Hall voltage using Maxwell's Equations and Coulomb's law for magnetic poles, and implemented by using finite element methods such as those of Pardo \emph{et al.} \cite{ChenIEEE1991,PardoIEEE2004,PardoIEEE2005,PardoJAP2002}.  Second, a calibration of the absolute value of the susceptibility must be obtained by comparing with measurements on a bulk sample for which the intrinsic susceptibility can be easily measured or calculated.  In our case, for example, we used SQUID-based measurements in zero  transverse field of the $\chi$ obtained from bulk measurements as a function of aspect ratio extrapolated to the value for an infinitely long sample, as described earlier \cite{ShiqiPRB2010}.  These steps provide a relationship that allows us to convert between the $\chi$ and the $V_\mathrm{Hall}$.

The relationship can then be used to correct for the demagnetizing effect by extension where appropriate. In our case, the parameters and the relationships that were obtained by a simple analysis using electromagnetic theory in the absence of transverse field were then assumed to be unaltered in the presence of transverse magnetic field, enabling us to correct (by extension) for the demagnetization effect (to obtain $\chi$ from $V_\mathrm{Hall}$) in a field. We note that the results obtained by this procedure agree with the MFA calculations of Millis {\it et al.} \cite{MillisPRB2010}.

\section{Acknowledgments}
We thank Dimitry Garanin and Eugene Chudnovsky for insightful discussions and George Christou for the Mn$_{12}$-ac samples that yielded the data on which this paper is based.  MPS acknowledges support from ARO W911NF-13-1-0125. AJM acknowledges support of NSF-DMR-1006282. YY acknowledges support by the Israel Science Foundation Grant No.164/12.  EP acknowledges the financial support of the European Commission for the project NMP3-LA-2012-280432 ``EUROTAPES" and from the Slovak Research and Development Agency under contract No. DO7RP-0003-12.  ADK acknowledges support from NSF-1309202.

\appendix
\section{Critique of Assumption of Uniform Magnetization}
\label{critiqueAppendix}
The assumption of uniform magnetization is commonly made in magnetic measurements. In this section, we will show that it is not suitable for local magnetometry.

With the coordinates defined in Fig.~\ref{setupSingle}, $\mathbf{M}$ is uniform means $\mathbf{M}=M_y\hat{y}+M_z\hat{z}$ inside the sample (we have applied field in both $y$ and $z$ directions). We have $\triangledown'\cdot\mathbf{M}(\mathbf{r}')=0$, so that the only contribution to $\Phi_M$ comes from the surface pole density (Eq. \ref{PhiMintegral1}):
\begin{align*}
\Phi_M(\mathbf{r})&=\frac{1}{4\pi}\int_{-x_0}^{x_0}\int_{-z_0}^{z_0}(\frac{-M_y}{|(x,y,z)-(x',-y_0,z')|}\\&+\frac{M_y}{|(x,y,z)-(x',y_0,z')|})dx'dz'\\
&+\frac{1}{4\pi}\int_{-x_0}^{x_0}\int_{-y_0}^{y_0}(\frac{-M_z}{|(x,y,z)-(x',y',-z_0)|}\\&+\frac{M_z}{|(x,y,z)-(x',y',z_0)|})dx'dy'\\
\end{align*}
\begin{align*}
\frac{1}{\mu_0}B_x&=H_x=-\frac{\partial\Phi_M}{\partial x}\\
&=\frac{1}{4\pi}\int_{-x_0}^{x_0}\int_{-z_0}^{z_0}(\frac{x-x'}{|(x,y,z)-(x',-y_0,z')|^3}\\&-\frac{x-x'}{|(x,y,z)-(x',y_0,z')|^3})dx'dz'\cdot M_y\\
&+\frac{1}{4\pi}\int_{-x_0}^{x_0}\int_{-y_0}^{y_0}(\frac{x-x'}{|(x,y,z)-(x',y',-z_0)|^3}\\&-\frac{x-x'}{|(x,y,z)-(x',y',z_0)|^3})dx'dy'\cdot M_z\\
&=K_1(\mathbf{r})\cdot M_y(H_y)+K_2(\mathbf{r}) \cdot M_z(H_z)
\end{align*}
where $K_1(\mathbf{r})$ and $K_2(\mathbf{r})$ are geometric factors depending only on the location of the sensor, and we assumed that the applied magnetic field is along $z$ direction.

From this, we can deduce a commonly used approximation in Hall sensor measurements: the fringing field $B_x$ is a linear function of $M_z$.  As we show below, if the Hall sensor is placed in the center of the sample along the $y$-direction ($y=0$), the $M_y$ term vanishes after averaging over the Hall area:
\begin{align*}
K_1(\mathbf{r})&=\frac{1}{4\pi}\int_{-x_0}^{x_0}\int_{-z_0}^{z_0}(\frac{x-x'}{|(x,0,z)-(x',-y_0,z')|^3}\\&-\frac{x-x'}{|(x,0,z)-(x',y_0,z')|^3})dx'dz'\\
&=\frac{1}{4\pi}\int_{-x_0}^{x_0}\int_{-z_0}^{z_0}(\frac{x-x'}{((x-x')^2+y_0^2+(z-z')^2)^{3/2}}\\&-\frac{x-x'}{((x-x')^2+y_0^2+(z-z')^2)^{3/2}})dx'dz'\\
&=0
\end{align*}
Thus, if the Hall sensor is on the middle line at $y=0$, then $B_x$ is proportional to only $M_z$ because $K_1$ is zero.  Note, however that $\chi_{zz}$ is related only to $M_z$ whether or not the Hall sensor is placed at $y=0$, since the xyz axis are aligned with the principal axis of the susceptibility tensor.

\begin{align*}
\chi_{zz}=\frac{\partial M_z}{\partial H_a}|_{H_a=0}=\frac{\partial(B_x/\mu_0-K_1\cdot M_y)/K_2}{\partial H_a}=\frac{1}{\mu_0 K_2}\frac{\partial B_x}{\partial H_a}
\end{align*}

The $\chi_{zz}$ measured at different $z$ positions of the sample should yield results that can be normalized by only multiplicative factors (a simple calibration of the sensor signal strength), i.e., the (1/$\chi_{zz}$) plotted as a function of temperature should all intercept at the same point on the temperature axis to give the same Curie Weiss temperature.

\begin{figure}[htbp]
\centering
\includegraphics[width=0.8\linewidth]{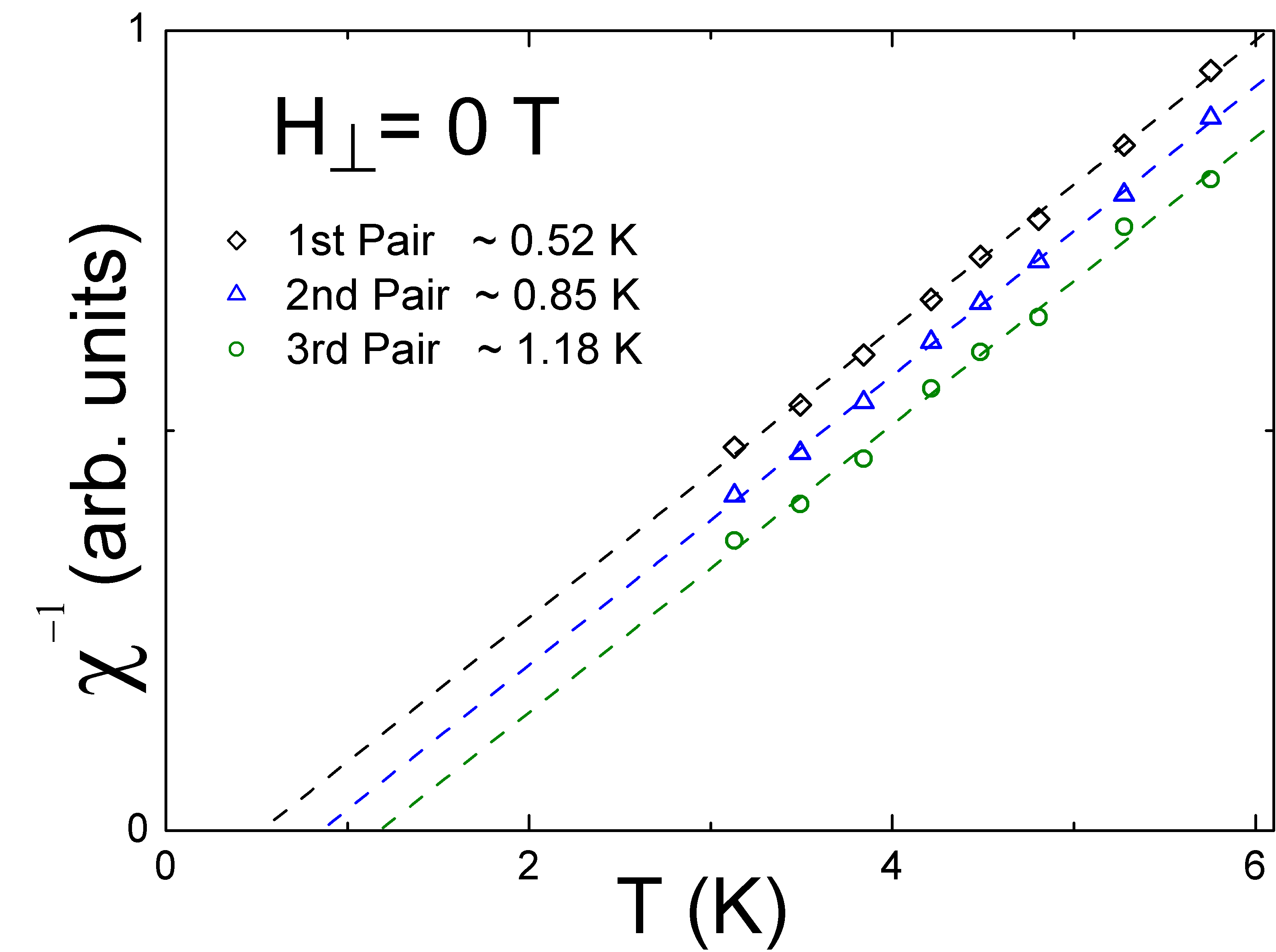}
\caption{Apparent susceptibility versus temperature at zero transverse field for the three locations defined in Fig. \ref{TcwComparison}. The values are normalized by each location's signal strength. The corresponding temperature intercepts, $T_W$, are noted in the legend.} \label{DiffProbes}
\end{figure}

First, this clearly disagrees with the experimental observation. Fig. \ref{DiffProbes} shows how the temperature intercept of the apparent inverse susceptibility $H_a/V_\mathrm{Hall}$ varies along the sample in zero transverse magnetic field. Fig. \ref{TcwComparison} showed same information in non-zero field. In addition to normalizing the values of the susceptibility, it is necessary to shift the curves along the T axis to achieve coincidence.

Second, we know from magnetostatics that the magnetization of a non-ellipsoidal sample is non-uniform, and depends on both $\chi$ and sample shape \cite{brownBook,ChikazumiBook}.

In contrast with global magnetization measurements, where large local variations in demagnetization field tend to average out, local magnetometry is sensitive to the local variations, thereby complicating the interpretation of the measurements.  On the other hand, local sensors provide an opportunity to study the detailed magnetic response with good spatial resolution.

\section{Magnetic field averaged over a rectangular surface}
\label{finiteElementAppendix}
This section is a revised version of Appendix A of Pardo \textit{et al.} Ref.~\cite{PardoIEEE2004} with the typos corrected and equations reformatted. Noted that these typos are only misprints. The calculation in Ref.~\cite{PardoIEEE2004} used the correct equations.

We derive the magnetic field generated by a rectangular plate
with uniform surface pole density $\sigma$ and averaged over a rectangular
surface using the magnetic Coulomb law
\begin{align}
  \mathbf{H(r)}=\frac{1}{4\pi}\int_{S'}\frac{\sigma(\mathbf r')(\mathbf r-\mathbf r')}{|\mathbf r-\mathbf r'|^3}dS'
\end{align}

We will only consider the field component perpendicular to the
surface over which the average is made, according to the needs
in Eq. \ref{Ddef}. We call the magnetic field generated by the plate
$\mathbf H^{(\alpha)}$ with $\alpha=x$, $y$ or $z$ for the plate to be perpendicular to
the $x$, $y$ or $z$ direction, respectively. The corresponding average
over a rectangular surface is $\langle \mathbf H^{(\alpha)}\rangle_\beta$ with $\beta=x$, $y$,
or $z$ for the surface to be perpendicular to the $x$, $y$ or $z$ direction,
respectively.

$\mathbf H^{(x)}$, $\mathbf H^{(y)}$, and $\mathbf H^{(z)}$ may be calculated by direct integration
of the fields produced by point poles. Assuming the plate to be
centered at the origin with dimensions $2a(y)\times2b(z)$, $2a(z)\times2b(x)$, and
$2a(x)\times2b(y)$ for the cases of $\mathbf H^{(x)}$, $\mathbf H^{(y)}$, and $\mathbf H^{(z)}$ respectively, we
obtain:
\begin{align}
   \mathbf H^{(x)} = {}& \frac{\sigma}{4\pi\mu_0}[F_1(y,z,x;a,b)\mathbf{j}\nonumber\\&+F_1(z,y,x;b,a)\mathbf{k}+F_2(y,z,x;a,b)\mathbf{i}]\\
   \mathbf H^{(y)} = {}& \frac{\sigma}{4\pi\mu_0}[F_1(z,x,y;a,b)\mathbf{k}\nonumber\\&+F_1(x,z,y;b,a)\mathbf{i}+F_2(z,x,y;a,b)\mathbf{j}]\\
   \mathbf H^{(z)} = {}& \frac{\sigma}{4\pi\mu_0}[F_1(x,y,z;a,b)\mathbf{i}\nonumber\\&+F_1(y,x,z;b,a)\mathbf{j}+F_2(x,y,z;a,b)\mathbf{k}]
\end{align}
where functions $F_1(u,v,w;t,d)$ and $F_2(u,v,w;t,d)$ are defined as:  \begin{align}
   &F_1(u,v,w;t,d)=\nonumber\\
   &+f_1(u+t,v-d,w)-f_1(u+t,v+d,w)\nonumber\\&+f_1(u-t,v+d,w)-f_1(u-t,v-d,w)\\
   &F_2(u,v,w;t,d)=\nonumber\\
   &-f_2(u+t,v-d,w)+f_2(u+t,v+d,w)\nonumber\\&-f_2(u-t,v+d,w)+f_2(u-t,v-d,w)
\end{align}
and $f_1(u',v',w')$ and $f_2(u',v',w')$ are given by
\begin{align}
   f_1(u',v',w')=&\mathrm{arcsinh}\frac{v'}{\sqrt{u'^2+w'^2}}\\
   f_2(u',v',w')=&\arctan\frac{u'v'}{w'\sqrt{u'^2+v'^2+w'^2}}
\end{align}
Note that $F_2(z,x,y;a,b)\mathbf{j}=F_2(x,z,y;b,a)\mathbf{j}$. $\mathbf H^{(y)}$ in the original paper used the second one..  Also note that In the paper, $F_2$ has a typo: $+f_2(u-t,v+d,w)$ should be $-f_2(u-t,v+d,w)$ or $f_2(t-u,v+d,w)$.

Once the field distribution is known, its average over a rectangular surface may be deduced by surface integration again. Assuming the rectangular surface to be centered at $(x_0,y_0,z_0)$ with dimensions $2a'(y)\times2b'(z)$, $2a'(z)\times2b'(x)$, and $2a'(x)\times2b'(y)$ for the cases of $\langle \mathbf H^{(\alpha)}\rangle_x$, $\langle \mathbf H^{(\alpha)}\rangle_y$, and $\langle \mathbf H^{(\alpha)}\rangle_z$ (which is just the numerical value of the matrix element in Eq. \ref{Ddef}), respectively, we obtain \begin{align}
   \langle H^{(x)}_x\rangle_x=&\frac{\sigma}{16\mu_0\pi a'b'}G_2(y_0,z_0,x_0;a,b,a',b')\\
   \langle H^{(x)}_y\rangle_y=&\frac{\sigma}{16\mu_0\pi a'b'}G_1(y_0,z_0,x_0;a,b,a',b')\\
   \langle H^{(x)}_z\rangle_z=&\frac{\sigma}{16\mu_0\pi a'b'}G_1(z_0,y_0,x_0;b,a,b',a')\\
   \langle H^{(y)}_x\rangle_x=&\frac{\sigma}{16\mu_0\pi a'b'}G_1(x_0,z_0,y_0;b,a,b',a')\\
   \langle H^{(y)}_y\rangle_y=&\frac{\sigma}{16\mu_0\pi a'b'}G_2(z_0,x_0,y_0;a,b,a',b')\\
   \langle H^{(y)}_z\rangle_z=&\frac{\sigma}{16\mu_0\pi a'b'}G_1(z_0,x_0,y_0;a,b,a',b')\\
   \langle H^{(z)}_x\rangle_x=&\frac{\sigma}{16\mu_0\pi a'b'}G_1(x_0,y_0,z_0;a,b,a',b')\\
   \langle H^{(z)}_y\rangle_y=&\frac{\sigma}{16\mu_0\pi a'b'}G_1(y_0,x_0,z_0;b,a,b',a')\\
   \langle H^{(z)}_z\rangle_z=&\frac{\sigma}{16\mu_0\pi a'b'}G_2(x_0,y_0,z_0;a,b,a',b')
\end{align}
where:
\begin{align}
   &G_1(u,v,w;t_1,d_1,t_2,d_2)=\nonumber\\&g_1(u,v+t_2,w+d_2;t_1,d_1)-g_1(u,v-t_2,w+d_2,t_1,d_1)\nonumber\\
   &-g_1(u,v+t_2,w-d_2,t_1,d_1)+g_1(u,v-t_2,w-d_2,t_1,d_1)\\
   &G_2(u,v,w;t_1,d_1,t_2,d_2)=\nonumber\\&g_2(u+t_2,v+d_2,w;t_1,d_1)-g_2(u+t_2,v-d_2,w;t_1,d_1)\nonumber\\
   &-g_2(u-t_2,v+d_2,w;t_1,d_1)+g_2(u-t_2,v-d_2,w;t_1,d_1)
\end{align}
with functions $g_1(u',v',w';t',d')$ and $g_2(u',v',w';t',d')$ defined as
\begin{align}
   g_1(u',v',w';t',d')=&+\tilde{f}_1(u'+t',v'-d',w')\nonumber\\&-\tilde{f}_1(u'+t',v'+d',w')\nonumber\\&+\tilde{f}_1(u'-t',v'+d',w')\nonumber\\&-\tilde{f}_1(u'-t',v'-d',w')\\
   g_2(u',v',w';t',d')=&-\tilde{f}_2(u'+t',v'-d',w')\nonumber\\&+\tilde{f}_2(u'+t',v'+d',w')\nonumber\\&-\tilde{f}_2(u'-t',v'+d',w')\nonumber\\&+\tilde{f}_2(u'-t',v'-d',w')
\end{align}
with
\begin{align}
   \tilde f_1(u'',v'',w'')=&-u''v''\arctan\frac{v''w''}{u''\sqrt{u''^2+v''^2+w''^2}}\nonumber\\&-\frac{w''}{2}\sqrt{u''^2+v''^2+w''^2}
   \nonumber\\&+v''w''\mathrm{arcsinh}\frac{v''}{\sqrt{u''^2+w''^2}}\nonumber\\&+\frac{v''^2-u''^2}{2}\mathrm{arcsinh}\frac{w''}{\sqrt{u''^2+v''^2}}\\
   \tilde f_2(u'',v'',w'')=&u''v''\arctan\frac{u''v''}{w''\sqrt{u''^2+v''^2+w''^2}}\nonumber\\&-w''\sqrt{u''^2+v''^2+w''^2}
   \nonumber\\&+u''w''\mathrm{arcsinh}\frac{u''}{\sqrt{v''^2+w''^2}}\nonumber\\&+v''w''\mathrm{arcsinh}\frac{v''}{\sqrt{u''^2+w''^2}}
\end{align}

Note that $G_2(z_0,x_0,y_0;a,b,a',b')=G_2(x_0,z_0,y_0;b,a,b',a')$, in the reference \cite{PardoIEEE2004}, the $\langle H^{(y)}_y\rangle_y$ term is set equal to the later form.

\section{Anisotropy in Mn$_{12}\mathrm{-ac}$}
\label{anisotropyAppendix}
\begin{figure}[htbp]
\centering
\includegraphics[width=0.8\linewidth]{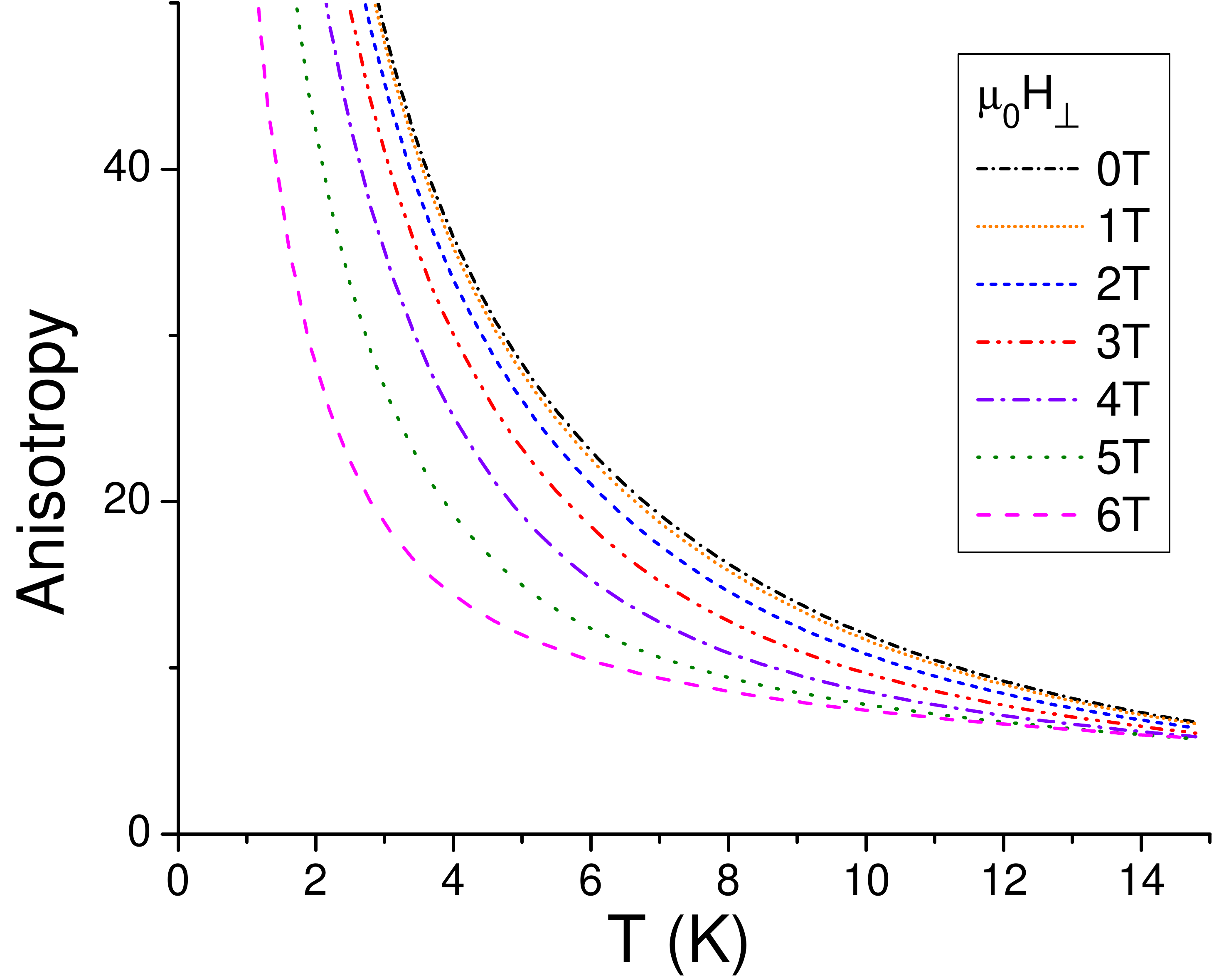}
\caption{Mn$_{12}$-ac single ion anisotropy {\it vs} temperature under different transverse fields} \label{anisotropy}
\end{figure}

Mn$_{12}$-acetate is a strongly anisotropic system with a c-axis in the easy direction, i.e., for the diagonal elements, $\chi_{zz}> \chi_{xx}=\chi_{yy}$. The single ion anisotropy, $\gamma=\chi_{zz}/\chi_{xx}=\frac{\Delta \langle S_z\rangle}{\Delta H_z}/\frac{\Delta \langle S_x\rangle}{\Delta H_\perp}$, can be estimated by the single ion Hamiltonian:
\begin{equation}
   \hat{\mathcal{H}}=-DS^2_z-BS^4_z-g\mu_BH_z S_z+g\mu_BH_\perp S_x
   \label{spinHam}
\end{equation}
where $D=0.548$ K, $B=0.0012$ K \cite{MillisPRB2010}. Figure \ref{anisotropy} shows the calculated $\gamma$ versus temperature for different $H_\perp$. In the following procedures, we assume that the single ion anisotropy is a good approximation of the crystal anisotropy.

\begin{figure}[htbp]
\centering
\includegraphics[width=0.8\linewidth]{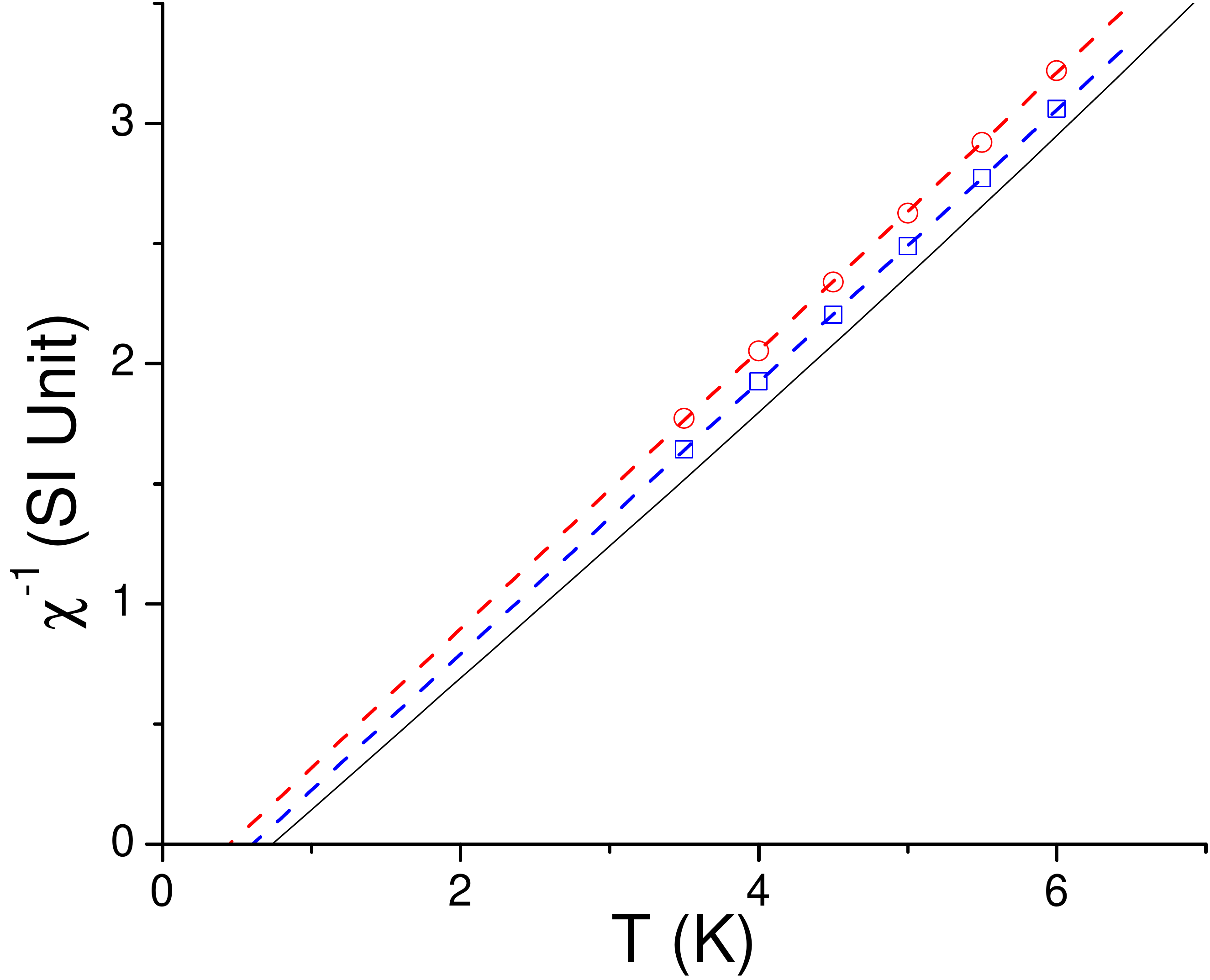}
\caption{MPMS measurement results: circles denote Mn$_{12}$-ac data of a crystal of aspect ratio of $c/a=1.5$; squares denote data of a crystal of aspect ratio of $c/a=3.3$. The dashed lines are linear fits to the data and the black line is the result of a mean-field.} \label{zeroField}
\end{figure}

In the paper of Shiqi \emph{et al.} \cite{ShiqiPRB2010}, the crystal longitudinal susceptibility $\chi_{zz}=\partial M_{z}/\partial H_z|_{H_z=0}$ was measured in a Quantum Design MPMS magnetometer for a set of crystals with different aspect ratio. The apparent Weiss temperature, $T_W$ obtained from the intercepts of $\chi^{-1}$ {\it vs} $T$ plotted against aspect ratio were extrapolated to infinite aspect ratio to obtain the intrinsic Weiss temperature for a very long, thin sample for which demagnetization effects are negligible.  Shiqi \emph{et al.} also found that the slope of the inverse susceptibility versus temperature between 3 K and 6 K is independent of sample aspect ratio. Using the demag-corrected $T_W$ and slope, we can normalize the parameter in the mean-field theory calculation by Millis \emph{et al.} \cite{MillisPRB2010} to obtain the temperature dependence of the susceptibility in the absence of a transverse field, as shown in Fig. \ref{zeroField} for the case of Mn$_{12}$-ac. A linear fit to the solid lines between 3 K and 6 K yields a slope of $\sim0.57$ and a temperature intercept about $\sim 0.85$ K, in agreement with the MPMS extrapolated value the for infinite aspect ratio sample.

With both anisotropy and $\chi_{zz}$ in hand, we have the full information for the susceptibility in zero transverse field:
 \begin{equation}
\mathbf{\mathbf{\chi}}=\left(\begin{array}{ccc}\chi_{xx} & 0 & 0 \\0 & \chi_{yy} & 0 \\0 & 0 & \chi_{zz}\end{array}\right)=\left(\begin{array}{ccc}1/\gamma & 0 & 0 \\0 & 1/\gamma & 0 \\0 & 0 & 1\end{array}\right)\chi_{zz},
\end{equation}
where $\chi_{ij}=\partial M_{i}/\partial H_j|_{H_j=0}$. By plugging this into Eq. \ref{sigmaeq}, we can include the anisotropy when applying this method to analyze the experiment data.

\end{document}